\documentclass{PoS}

\title{The Galactic Center as a Paradigm for \\
Low Luminosity Nuclei?\\
{\small What can be learned from SgrA* for the central engine and conditions of star formation
in nuclei of Seyfert galaxies and low luminosity nearby QSOs \\}
 {\small The K-band identification of the DSO/G2 source from VLT and Keck data }}

\ShortTitle{The Galactic Center}

\author{\speaker{Andreas Eckart}
\\
\\
        E-mail: \email{eckart@ph1.uni-koeln.de}}

\author{
S. Britzen$^{2}$, 
M. Horrobin$^{1}$,
M. Zamaninasab$^{2}$, 
K. Mu\v{z}i\'{c}$^{3}$, 
N. Sabha$^{1,2}$, 

B. Shahzamanian$^{1,2}$,
S. Yazici$^{1}$, 
L. Moser$^{1}$, 
J. Zuther$^{1}$, 
M. Garcia-Marin$^{1}$, 

M. Valencia-S.$^{1}$, 
M. Bursa$^{4}$, 
G. Karssen$^{1}$, 
V. Karas$^{4}$, 
B. Jalali$^{1}$, 
M. Vitale$^{2,1}$,

M. Bremer$^{1}$,
S. Fischer$^{1}$, 
S. Smajic$^{1}$, 
C. Rauch$^{2}$, 
D. Kunneriath$^{4}$, 
J. Moultaka$^{5,6}$,

C. Straubmeier$^{1}$,
Y.E., Rashed$^{1}$,
C. Iserlohe$^{1}$,
G. Busch$^{1}$,
K. Markakis$^{2,1}$, 

A. Borkar$^{2,1}$, 
A. Zensus$^{2,1}$
\\
1) I. Physikalisches Institut, Universit\"at zu K\"oln,
           Z\"ulpicher Str. 77,
           50937 K\"oln, Germany
\\
2)            Max-Planck-Institut f\"ur Radioastronomie, 
           Auf dem H\"ugel 69, 
	   53121 Bonn, Germany
\\
3)            European Southern Observatory, Alonso de Cordova 3107, Vitacura, Casilla 19, Santiago, 19001, Chile 
\\
4)            Astronomical Institute, Academy of Sciences, Bocni II 1401, CZ-141 31 Prague, Czech Republic
\\
5)            Universit\'e de Toulouse; UPS-OMP; IRAP; Toulouse, France
\\
6)            CNRS; IRAP; 14, avenue Edouard Belin, F-31400 Toulouse, France
          }

\abstract{
The super-massive 4 million solar mass black hole (SMBH) SgrA*
shows flare emission from the millimeter to the X-ray domain.
The nucleus of the Milky Way has properties 
(stellar cluster, young stars, molecular gas and an accreting SMBH)
that resemble those of currently higher luminous 
Low Luminosity Active Galactic Nuclei.
A detailed analysis of the infrared light curves shows that the flares are probably 
generated in a single-state process
forming a power-law distribution of the flux density.
Near-infrared polarimetry shows signatures of strong gravity that are 
statistically significant against randomly polarized red noise.
Details of the emission mechanism are discussed in a synchrotron/self-Compton model.
SgrA* also allows to study the interaction of the SMBH with the
immediate interstellar and gaseous environment of the central stellar cluster.
Through infrared imaging of the central few arcseconds 
it is possible to study both inflow and outflow phenomena linked to the 
SgrA* black hole.
In this context we also discuss the newly found dusty object that approaches SgrA*
and present a comparison between recent Keck and VLT K-band data that clearly
supports its detection as a $\sim$19$^m$ K'-band continuum source.
}

\FullConference{Nuclei of Seyfert galaxies and QSOs - Central engine \& conditions of star formation ,\\
		November 6-8, 2012\\
		Max-Planck-Insitut f\"ur Radioastronomie (MPIfR), Bonn,  			Germany}

\begin{document}

\section{Introduction}

Sagittarius~A* (SgrA*) at the center of our galaxy is a highly variable 
near-infrared (NIR) and X-ray source which is associated with a 
4$\times$10$^{6} M_{\odot}$ central SMBH.
Zamaninasab et al. (2010, 2011) has shown that the polarized NIR flares exhibit patterns 
of strong gravity, as expected from in-spiraling material very close to the 
black hole's horizon.
Therefore, it is mainly the strong flux variability 
that gives us certainty that we study the immediate vicinity of a SMBH
and investigate the largely unknown flare processes. For bright
high  signal to noise flares, relativistic modeling of NIR polarization data and 
SED modeling of the multi-wavelength data
can discriminate between 
pure Synchrotron and Self-Compton models (e.g. Eckart et al. 2012).

\begin{figure}
\begin{center}
\includegraphics[width=1.0\textwidth]{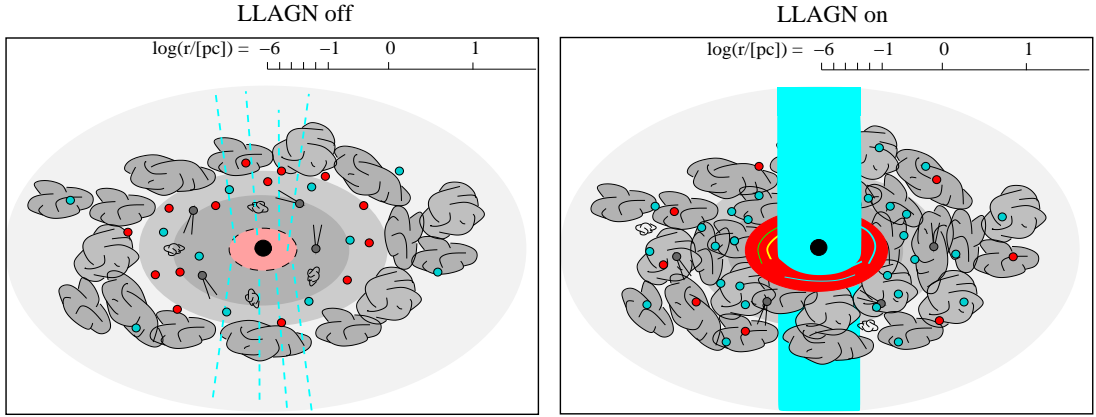}
\caption[]{Inclined view on an LLAGN in its 'off' and 'on' state. 
A logarithmic scale along the semi-major axis is given.
The molecular and atomic gas reservoir is indicated by dark clouds.
The stellar cusp is indicated by the grey background.
A temporary or truncated disk and a more substantial accretion disk
are depicted by the pink and red disks surrounding the central black hole.
The cyan lines and tube-like structure represent the 
not necessarily collimated wind or outflow off-state and a possible jet in the
on-state.
Red giants and blue young stars are shown as filled red and blue circles.
In the off state the activity is dominated by rare and sporadic accretion events
linked to dusty DSO-like objects (grey circles with commentary tails) and to smaller cloudlets 
that are clearly visible in the off-state image
(see sections 2.3, 2.4 \& 2.6
and Gillessen 2012, Eckart et al. 2013).
}
\label{fig0}
\end{center}
\end{figure}

Given the unambiguous traces of recent activity of the Galactic Center
it is worthwhile to compare its properties with those of
currently even more active sources such as the class of low 
Low Luminosity Active Galactic Nuclei (LLAGN).
A detailed comparison to the center of the
Milky Way and an in-depth discussion of the LLAGNs as a source class 
has been given by Contini (2011) and Ho (2008).
Contini (2011) finds that, based on radio as well as far-infrared continuum and line 
emission ([OI] 63$\mu$m \& 145$\mu$m, [NII] 122$\mu$m, [CII] 158$\mu$m),
the physical properties of the Galactic Center compare well to properties found for 
faint LLAGN.
Ho (2008) outlines that due to the short duty cycle of the Black Hole accretion, 
most AGN spend their life in a low state, such that the bulk of the population 
has relatively modest luminosities. 
He concludes that the absolute luminosity can no longer be used as 
a defining metric of nuclear activity, and a broader range of alternative 
mechanisms needs to be considered to explain the nature of LLAGNs.
Although a comparison of radio to X-ray continuum spectra of the Galactic Center 
to those of some bona fide LLAGNs 
(like NGC 3147, NGC 4579, NGC 4203, NGC 4168, NGC 4235, and NGC 4450)
shows remarkable similarities, the detailed e.g. radio properties of 
extreme low luminosity objects (like the GC) may be different from those of
higher luminosity sources (e.g.  Zuther et al. 2012, 2008).
Observational constraints and theoretical models suggest that the LLAGN are 
linked to radiation inefficient accretion flows (RIAF) with luminosities of 
10$^{-7}$$L_{Edd}$ or less,
and that at $L_{bol}$/ $L_{Edd}$$\le$10$^{-3}$ the BLR disappears and a 
truncated or only temporarily existing disk is very likely 
(Laor 2003, Nicastro 2000, Xu \& Cao 2007).
In Fig.\ref{fig0} we summarize these findings in a model of a LLAGN 
in its 'off'- and 'on'-state.
A more detailed understanding of LLAGN through a comparison to the Galactic Center 
will also be fruitful for the investigations of the fueling of central black holes and
nuclear star formation on more luminous nearby AGN
(e.g.  Smaji\'c et al. 2012, Fischer et al. 2012, Bremer et al. 2012, 
Valencia-S. et al. 2012, Scharw\"achter et al. 2011, Vitale et al. 2012).  

\section{Building Blocks of a Nucleus}
\label{buildingblocks}
Based on the findings presented by Contini (2011) and Ho (2008), the Galactic Center 
is ideally suited to study the building blocks (see sections \ref{cluster} to \ref{variability})
that are required to explain the thermal and non-thermal activities in a LLAGN.
Given the correlation between the black hole mass and the stellar velocity dispersion in the bulge 
(which apparently largely holds independently on the possible presence a pseudo-bulge; Beifiori et al. 2012)
and the fact that LLAGN are found in a large fraction of all massive galaxies (Maoz 2008),
we can take the properties of the Galactic Center stellar cluster as a paradigm for bulges hosting the LLAGN.
On size scales of the nuclear bulge the efficiency with which stars are formed from 
molecular gas appears to be dependent on the bulge properties (Fisher et al. 2013).
However, the non-thermal AGN activity does not appear to be strongly linked to the amount of
molecular gas available within the overall bulge 
(e.g.  Krips et al. 2011, Bertram et al. 2007, Garcia-Burillo et al. 2008).
Based on the NUGA sample of nearby AGN, Garcia-Burillo \& Combes (2012; see also Casasola et al. 2010, 2011)
show that secular evolution and dynamical decoupling of molecular gas structures are more likely the
key ingredients to understand the AGN fueling and hence the overall AGN activity.
Therefore, we speculate that also the star formation properties in the immediate vicinity ($<$ a few parsec)
of the SMBH will depend on these factors rather than on more global bulge properties.
This makes the Galactic Center an ideal case to study the conditions and impact of star formation for LLAGN
in the immediate nuclear region.

\subsection{The Stellar Cluster}
\label{cluster}
The investigation of stellar properties 
of the Galactic Center
(proper motions, radial velocities, colors, spectra, and variability)
gives a clear view on the structure, dynamics, and stellar populations 
of the central cluster.
First infrared stellar proper motion measurements were described by
Eckart \& Genzel (1996, 1997).
Eckart et al. (2002) showed that the star S2 is in fact on a bound orbit and gave first orbital 
elements. More detailed elements of S2 and other stars were then successfully 
published by Sch\"odel et al. (2002, 2003) and Ghez et al. (2003).
Further improved orbital elements and derivations of the distance to the center from
stellar orbits and proper motions were given by Eisenhauer et al. (2005) and Gillessen et al. (2009).
Sch\"odel, Merritt \& Eckart (2009) measured the proper motions of  more than 
6000 stars within the central stellar cluster (see also Sch\"odel et al. 2010).
These stars are within about 1.0 pc of SgrA*. They find that the bulk 
of the cluster rotates parallel to the Galactic rotation, 
while the velocity dispersion $\sigma$ appears isotropic (see details in their paper). 
A Keplerian fall-off of $\sigma$ due to the central 
point mass is detectable at separations of only a few 0.1~pc 
from SgrA* (see also Sabha et al. 2012).
The authors find a best-fit black hole mass 
of 3.6$\pm$(0.2-0.4)$\times$10$^6$M$\odot$ which is consistent with the
mass inferred from the orbits of the individual S-cluster stars close to SgrA*. 
Hence, in terms of the SMBH mass the Galactic Center may be taken as a 
representative case for other nuclei, since comparable or even lower black hole 
masses are also indicated in other AGN (e.g. Valencia-S. et al. 2012).

\subsection{The conditions for Star Formation}
\label{starformation}
The polarization of stars measured at NIR wavelengths 
(Witzel et al. 2011; Buchholz et al. 2011, 2013, Rauch et al. 2013) allows us to separate 
objects that are polarized due to their local or intrinsic dust 
distributions from those that are polarized (predominantly only) due to the Galactic foreground.
The detailed near- and mid-infrared properties of these dusty objects are outlined in 
several papers (Rauch et al. 2013, Eckart et al. 2013, Moultaka et al. 2009, Moultaka et al. 2004, 
Viehmann et al. 2006, 2005).
The Galactic Center stellar cluster contains several young luminous
He-stars that arrange themselves in at least one disk, indicating that they have been 
formed recently
(Do et al. 2009, 2003, Eisenhauer et al. 2005, Levin \& Beloborodov, 2003, and references there in).
The IRS13N complex (Eckart et al. 2004, Muzic et al. 2008)
consists of a group of infrared excess sources located just north of IRS13E .
Together with the 
cometary shaped objects X3 and X7 (Muzic et al. 2010) and the G2/DSO 
(Gillessen 2012, Eckart et al. 2013), these are clear indications for the presence of 
dust enshrouded objects and ongoing star formation in the central stellar cluster.
The Galactic Center can therefore serve as a laboratory where the conditions
for star formation in a dense nuclear stellar cluster can be studied in detail.

\begin{figure}
\begin{center}
\includegraphics[width=.6\textwidth]{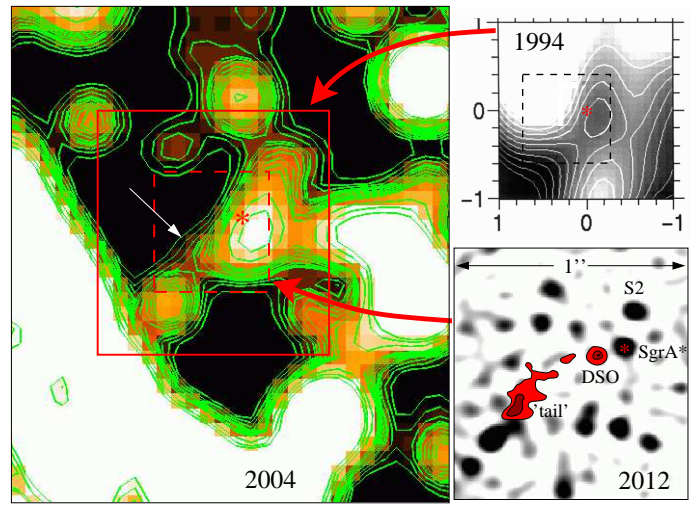} 
\caption[]{The 2004 VISIR and 1994 Palomar 
(Stolovy et al.  1996) 8.6 micron images, a source/flux ridge 
at the position of the DSO/G2 'tail' can be identified (white arrow). 
A 2012 NACO Ks-band image with the DSO 'tail' contours in red is shown. 
The image scales are given in arcsecond.}
\label{fig1}
\end{center}
\end{figure}

\begin{figure}
\begin{center}
\includegraphics[width=1.0\textwidth]{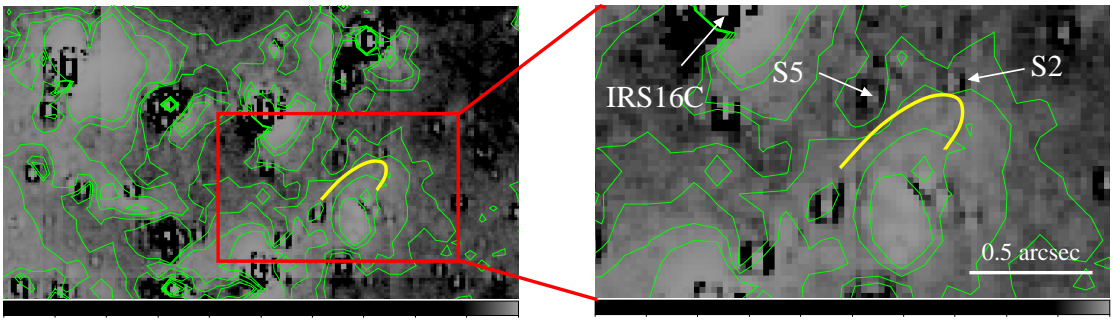}
\caption[]{
The Br$\gamma$ line emission (grey scale and contours) towards the central few
arcseconds as obtained from narrow band filter observations with NACO (Kunneriath et al. 2012).
The positions of some stars are labeled in the continuum subtracted image.
We also show the approximate orbit of the DSO from 2002 to beyond 2013 (Eckart et al. 2013).
The entire region is highly confused in its Br$\gamma$ line emission.
}
\label{fig3}
\end{center}
\end{figure}

\subsection{The Dusty S-cluster Object}
\label{DSO}
Recently, Gillessen et al. (2012, 2013) reported a fast moving 
infrared excess source (G2) which they interpret as a core-less gas and dust cloud
approaching SgrA* on an elliptical orbit.
Eckart et al. (2013) present first K$_s$-band identifications and proper 
motions of this dusty S-cluster object (DSO). 
In 2002-2007 it is confused with star S63, but free of confusion again since 2007. 
Its NIR colors and a comparison to other sources in the field 
imply that it could rather be an IR excess star than a 
core-less gas and dust cloud (Eckart et al. 2013).
At a temperature of 450~K a pure dust contribution is very unlikely (Fig.15 in Eckart et al. 2013).
Also we find very compact L'-band emission (<0.1'') contrasted by the 
extended ($\sim$0.2'')  Br$\gamma$ emission reported by 
Gillessen et al. (2012, 2013) and modeled by 
Schartmann  et al. (2012) and Burkert  et al. (2012).
The presence of a star will change the expected accretion phenomena,
since a stellar Roche lobe may retain much of the material (Eckart et al.  2013)
during and after the peri-bothron passage.

From Fig.\ref{fig1} we can see that in 2004 there was a 8.6$\mu$m 
source component at the position of the DSO in 
2004 or the position of the DSO/G2 'tail' in 2012 (Gillessen  et al. 2012, 2013).
Extended flux density at that location was already indicated 
in the MIR continuum map presented by Stolovy et al. (1996).
Given the 8.6$\mu$m dust emission at the current 'tail' position 
(Fig.\ref{fig1}) and the large amount of confusion emission in the central region
(Fig.\ref{fig3}) this suggests another interpretation:
The  DSO/G2 'tail' seen in recent years is in fact a 
background component, possibly within the mini-spiral gas/dust flow. 
This is supported by the very low radial velocity and
proper motion of the extended 'tail' component 
(Fig.7 in Gillessen et al. 2013 and Fig.10 in Schartmann et al. 2012).
If the 'tail' component is indeed a background source that is not associated to the
fast moving dusty object, then the DSO may be a
compact source comparable to the cometary shape sources X3 and X7
(Muzic et al. 2010). Its smaller size compared to X3 and X7 can be explained
by the higher particle density within the accretion stream close to 
SgrA* (e.g. Shcherbakov \&  Baganoff 2010).
Its size also depends on how possible earlier passages close to SgrA*
influence the distribution of gas and dust close to a possible 
star at the center of the DSO.
The long dust lane feature that crosses the mini-spiral to the south-east of the 
DSO (Fig.7 in Gillessen et al. 2013) may then simply be a consequence of the 
interaction between the two rotation gas disk 
components that are associated with the northern arm and the eastern arm
(Fig.21 in Zhao et al. 2009, and Fig.10 in Vollmer \& Duschl 2000).
If upcoming observations can confirm that the 
'tail' component is not associated with the DSO, 
it does not need to be taken into account by future simulations.
New simulations are needed that include the presence of a star
and go beyond the assumption of a pure gas/dust cloud.

\begin{figure}
\begin{center}
\includegraphics[width=0.9\textwidth]{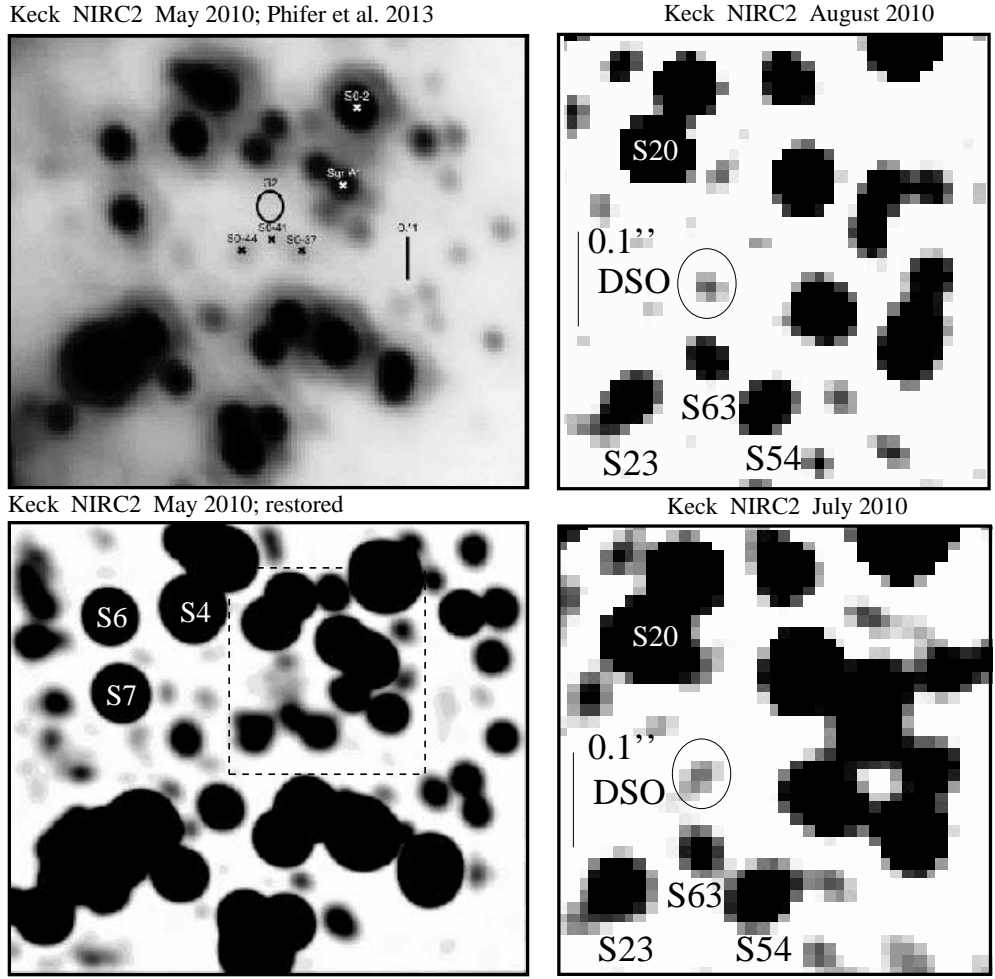}
\caption[]{
Comparison between Fig.2c by Phifer et al. (2013) (top left) 
and the restored version (bottom left) showing
fainter structure in the same image that corresponds surprisingly well to structure 
reported by Eckart et al. (2013) in Fig.A.1.
The public 2010 August 2010 (top right) and July (bottom right) Keck data reduced.
The DSO can clearly be identified close to the center of the 3$\sigma$ 
contour of the Br$\gamma$ line emission reported by Phifer et al. (2013).
This contour line is also shown as a black ellipse in the panels on the
right side.
The dashed rectangle in the bottom left image depicts the area of the
image sections shown on the right.
}
\label{fig4}
\end{center}
\end{figure}

\begin{table}[htb]
\begin{center}
\begin{tabular}{llrlllll}
\hline 
\hline 
date          &decimal & integration & R.A.   &   DEC.  \\
              &date    & time (sec)  &offset  & offset  \\
              &        &             &arcsec  & arcsec  \\ \hline
2008 05 20    &2008.38 &   140       &0.179   & -0.062  \\
2009 05 02    &2009.33 &   616       &0.170   & -0.051  \\
2010 07 05/06 &2010.52 &  4760       &0.151   & -0.039  \\
2010 08 14/15 &2010.62 &  4760       &0.144   & -0.042  \\
2011 07 18    &2011.55 &  5236       &0.117   & -0.023  \\
\hline 
\hline 
\end{tabular}
\caption[]{
List of public Keck data used for the present analysis and positions of the 
K'-band identification  of the DSO based on this data. 
From their scattering we estimate the uncertainties of the R.A. and DEC. offsets to be 
of the same order as the value of 14~mas given by Eckart et al. (2013).
\label{tab01}
}
\end{center}
\end{table}

\subsection{Confirmation of the K-band continuum detection of the DSO through Keck data}
\label{Keck}

Given that the role of the DSO is important in the context of nuclear stellar populations 
and that its physical structure is essential for predictions 
and interpretations of the expected interaction with SgrA*, a comparison to
additional observational data at 2$\mu$m wavelength is useful.
Recently, Phifer et al. (2013) stated that, based on Keck data, they cannot detect the 
DSO in K'-band down to a limit of K=20. 
In a K'-band image they show an empty aperture at the position of the Br$\gamma$ 
emission of the DSO. Size and shape of the aperture are given by the 3$\sigma$ contour line
of the Br$\gamma$ line emission.
Indeed, a simple visual inspection of this aperture in Fig.2c in Phifer et al. (2013) 
shows no source fainter than about 18.5 magnitude. 
We took Fig.2c by Phifer et al. (2013), removed all labels within the image by interpolation
and reestablished a linear flux density scale using known source fluxes from the corresponding
VLT data.
The result is shown in Fig.\ref{fig4}.
It reveals a structure surprisingly similar to what
Eckart et al. (2013) show for 2010 and we can clearly identify flux density at the position 
of the DSO counterpart at K'-band.

\begin{figure}
\begin{center}
\includegraphics[width=0.85\textwidth]{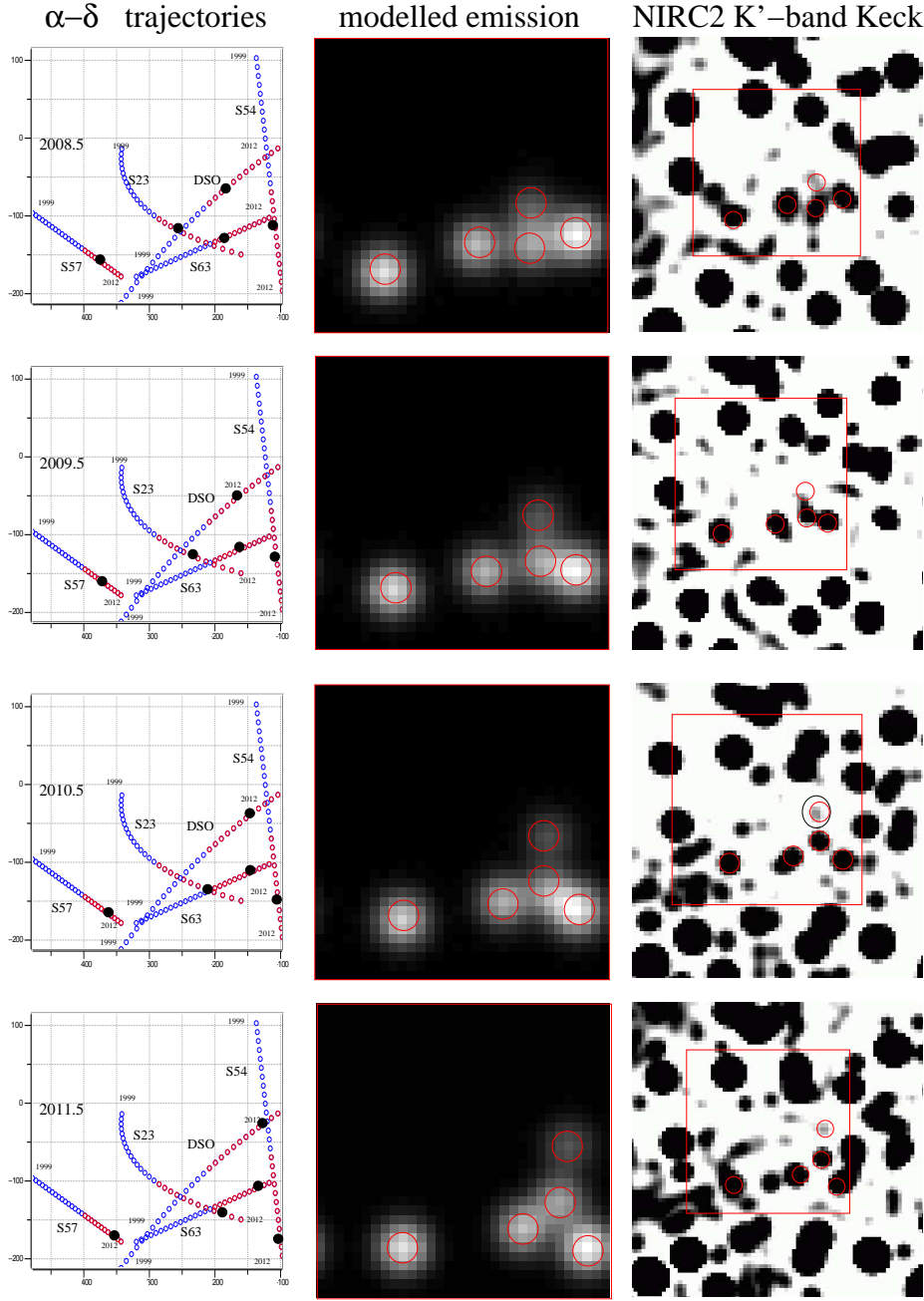}
\caption[]{ K'-band identification of the DSO using archival Keck NIRC2 
data for epochs 2008 till 2011
 and using the improved kinematic model from Tab~\ref{tab02}.
 We show the relative positions 
 of the stars S23, S57, S54, S63 and the DSO 
 over the years 2008 - 2011
 in comparison with K'-band Lucy-deconvolved Keck NIRC2 images (at $\sim$40mas resolution).
 We show the results of the model calculation (left) with the location of the sources on their sky-projected
 track (dot interval 0.5 years; 1999-2006.5 in blue, 2007-2012 in red),
 an image of the model with the sources indicated by red circles (middle; image size is 380mas$\times$380mas), and
 a deconvolved K'-band image with the modeled image section overlayed (right).
 For all years a source can clearly be 
 identified at the L'-band position of the DSO.
 In 2008 it emerges from the
 confusion with S63 and shows up between the stars S23, S63, and S54 
 (alias S0-44, S0-41, S0-37 in Keck nomenclature) just north of S63.  
 For 2010 we also show as a black ellipse the Br$\gamma$ 3$\sigma$ contour 
 for the DSO given by Phifer et al. (2013).}
\label{fig5}
\end{center}
\end{figure}

In order to further confirm the presence of the DSO at 2$\mu$m wavelength 
we reduced all publically available 
K'-band imaging data obtained by the Keck that contain the SgrA* region. 
The results for epochs 2008, 2009, 2010, and 2011 are shown in Fig.\ref{fig5} and 
listed in Tab.~\ref{tab01}.
In order to fully exploit the data, we applied a Lucy deconvolution 
algorithm using the brightest source (IRS16SW) in the frame as a PSF reference.
Inspection of other sources (e.g. IRS16C) in the field shows that 
the PSF structure at the location of the Airy ring changes on the 2\% level 
as a function of position.
In addition to a limited size PSF and a residual background
this results in brightness dependant differential artifacts
after deconvolution in the immediate surroundings of  bright sources S4, S6, and S7. 
In the right panels of Fig.\ref{fig4} we show the DSO in a region 
that is sufficiently far away from bright stars 
and is therefore not heavily affected by this effect at the given sensitivity.

\begin{figure}
\begin{center}
\includegraphics[width=0.85\textwidth]{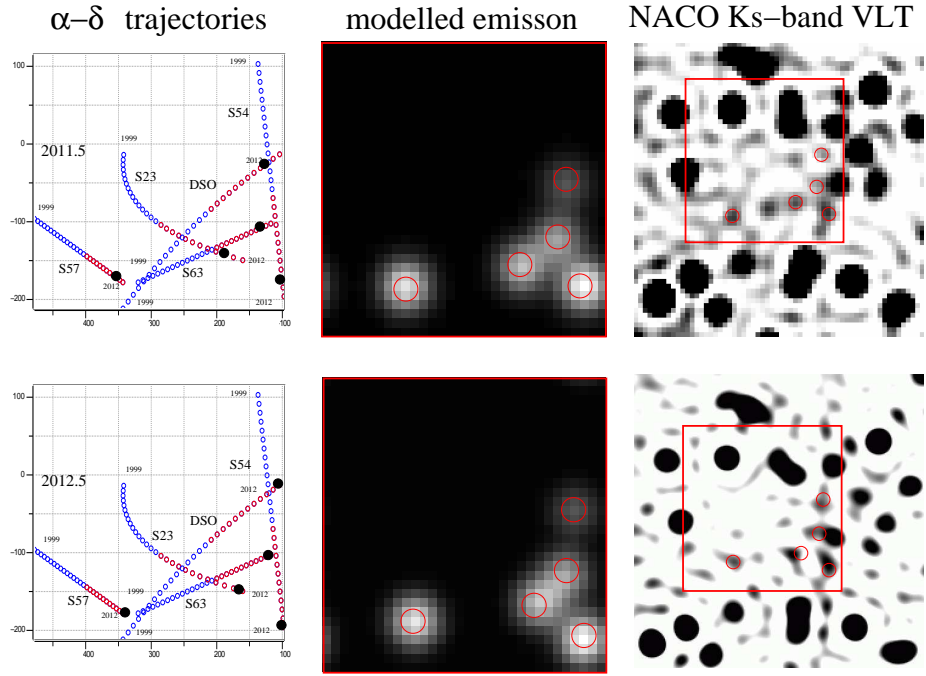}
 \caption[]{K$_s$-band identification of the DSO for epochs 2011 and  2012 shown 
 with deconvolved NACO data (at $\sim$40mas resolution)
 and using the improved kinematic model from Tab~2.
 See caption of Fig.\ref{fig5}.
 }
\label{fig55}
\end{center}
\end{figure}

Based on the 2011 and 2012 Keck and NACO data we improved our kinematic model for 
stars S23, S63, S54, and S57, such that their position is now better represented in 
the crowed region 100mas to 150mas south of SgrA*. For this we changed the model 
parameters listed in Tab.6 by Eckart et al. (2013) by less than 3 times the 1$\sigma$
uncertainty listed in the caption of that table. The improved parameters are listed
here in Tab.~\ref{tab02}.
We show the results for epoch 2011 and 2012 for the NACO data in comparison
with the new kinematic model in Fig.~\ref{fig55}.

\begin{figure}
\begin{center}
\includegraphics[width=1.0\textwidth]{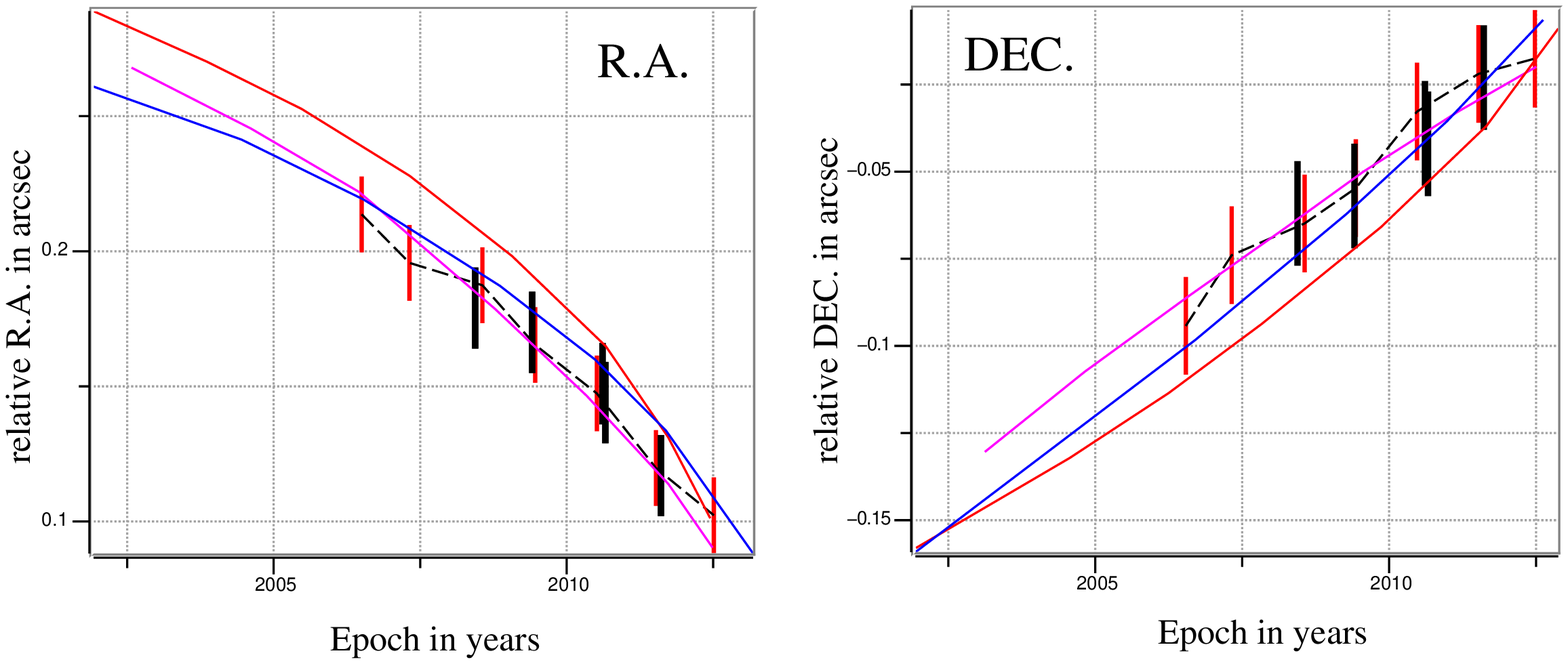}
\caption[]{
A comparison between the L'-band tracks of the DSO used by Gillessen et al. (2013) 
(magenta line; L'-band),
Phifer et al. (2013) (blue (K'-band Br$\gamma$) and red (L'-band) lines). 
We also show the coordinates obtained from the
K$_s$-band identification by Eckart et al. (2013) using VLT NACO data 
(data points with red error bars connected by a black dashed line).
Data point with thick black error bars represent the K'-band 
identification based on Keck NIRC2 data as presented here.
}
\label{fig6}
\end{center}
\end{figure}

For the Keck data shown in Figs.~\ref{fig4} and \ref{fig5} 
we find at all epochs very similar structures compared to those derived from the 
VLT data presented by Eckart et al. (2013).
Based on the comparison to VLT NACO L'- and K$_s$-band data (Eckart et al. 2013, Gillessen et al. 2013) 
we can clearly identify the DSO in its K'-band continuum emission as measured by the NIRC2 camera 
at the Keck telescope.
Using the relative astrometric reference base on the positions of SgrA*, S2  as well as the
positions of sources 
S23, S63, S54 (alias S0-44, S0-41, S0-37 in Keck nomenclature)
as derived from their modeled projected orbits (Eckart et al. 2013) 
we can include the Keck results in the plots showing the DSO's right ascension and 
declination as a function of time. A comparison between the L'-band and K-band Br$\gamma$ 
fits to the DSO trajectory published by Gillessen et al. (2013) and Phifer et al. (2013) 
as well as our K$_s$-band NACO (Eckart et al. 2013) and K'-band Keck identifications 
of the DSO is shown in Fig.\ref{fig6}.

The comparison demonstrates that within the uncertainty of 14~mas (Eckart et al. 2013, Fig.A6) 
our DSO identification is in good 
agreement with the L'- and K$_s$-band results by Gillessen et al. (2013) 
and Eckart et al. (2013) as well as the K-band Br$\gamma$ line emission 
results by Phifer et al. (2013).
With respect to the VLT L'- and K$_s$-band results,
the L'-band graph presented by Phifer et al. (2013) is systematically off by  
about 20 to 25~mas to the east and south. The comparison of the Keck and VLT results 
shows that there is a clear identification of the DSO in the K'-band continuum.
Despite the admirably small formal errors in positions and orbital fits,
the results appear to be affected by systematic effects.
Our estimate of the uncertainties of 14~mas apparently represents the current 
systematic and statistical uncertainties quite well.
This indicates that the combined L- and K-band results from both telescopes do 
currently not allow a very precise prediction of the DSO periapse.

Of course positions and flux densities for faint sources 
will not be as exact as for the bright members of the S-star cluster. 
The investigation of faint states of SgrA* by  Witzel et al. (2013) 
showed that the flux density information towards low fluxes 
becomes increasingly unreliable.
Sabha et al. (2012) showed that the flux densities and positions of 
faint sources will be influenced by the grainy and time variable 
distribution of faint cluster members in the background.
Hence, one cannot expect that the information on flux density 
and position of the DSO is as accurate as it may possibly be for
the brighter S-cluster sources.

\begin{table}[htb]
\begin{center}
\begin{tabular}{lrrrrrrrr}\hline \hline
Name &
Epoch &
Flux &
$\Delta$$\alpha$ &
$\Delta$$\delta$ &
pm$_{\alpha}$~~~ &
pm$_{\delta}$~~~ &
acc$_{\alpha}$~~~ &
acc$_{\delta}$~~~ \\
     &
year &
ratio &
mas &
mas &
mas yr$^{-1}$ &
mas yr$^{-1}$ &
mas yr$^{-2}$ &
mas yr$^{-2}$ \\ \hline
S23 & 2005.47&  1.01  & 307.4&   -89.1&  12.81&   10.17&  -1.153&   0.225\\
S57 & 2007.46&  1.24  & 393.6&  -147.4&   9.99&    6.05&   0.000&   0.000\\
S63 & 2004.00&  0.83  & 245.0&  -150.0&  15.00&   -5.60&   0.000&   0.000\\
DSO & 2002.00&  0.30  & 295.0&  -160.0&  16.50&  -16.40&  -0.160&  -0.240\\
S54 & 2006.59&  1.36  & 115.4&   -60.1&   3.05&   23.80&   0.000&   0.000\\
\hline \hline
\end{tabular}
\caption[]{Improved kinematic modeling of the proper motions of the DSO and nearby sources.
Epochs, flux ratios, positions, proper motions, and accelerations used to simulate the DSO and
the neighboring stars with results shown 
here in Figs.~\ref{fig5} \& \ref{fig55} and in
Figs. A1, A2 \& A3 by Eckart et al. (2013).
The flux ratios have been calculated with respect to the mean flux derived (m$_K$=17.6) from the K$_s$-band 
magnitudes of S23, S54, S57, and S63 as given by Gillessen et al. (1999). 
They correspond (with an estimated uncertainty of 30\%) to the
relative fluxes suggested by the modeling presented in the figures listed above.
From the quality of the fit in these figures we estimate that the uncertainties 
are 10 mas for the coordinates, 1 mas~yr$^{-1}$ for the proper motions,
and  0.1 mas~yr$^{-2}$ for the accelerations.
}
\label{tab02}
\end{center}
\end{table}

\subsection{Molecular Gas and Dust}
\label{dustgas}
A recent LABOCA sub-mm map of the thermal GC emission surrounding SgrA*
is given by Garcia-Marin et al. (2011).
They find that the innermost tens of parsecs of our Galaxy are characterized 
by the presence of molecular cloud complexes surrounding SgrA*.
Using sub-mm maps, they describe the complex morphology of the molecular clouds and the 
circum-nuclear disk, along with their masses (of order 10$^5$ to 10$^6$~M$_{\odot}$), 
and derive also the temperature and spectral index maps of the regions under study. 
The authors conclude that the average temperature of the dust is 14$\pm$4~K. 
The spectral index map shows that the sub-mm emission of the 20 and 50 km/s 
clouds is dominated by dust emission. 
Comparatively, in the CND and its surroundings the spectral indices decrease 
toward SgrA* and has values between about 1 and -0.6. These numbers are mostly 
explained by a combination of dust, synchrotron, and free-free emission 
(in different proportions) at different positions. 
The presence of non-thermal emission also accounts for the 
apparent low temperatures derived in these very central regions. This indicates that
obtaining dust temperatures in this region is difficult and possibly misleading.
Requena-Torres et al. (2012) report sub-mm and millimeter spectroscopy measurements obtained 
with SOFIA/GREAT, Herschel/HIFI, and ground-based instruments.
The authors find that a superposition of various warm (200 to 500~K) gas phases is required
to explain the observations.
The densities around 10$^4$ to 10$^5$cm$^{-3}$ are derived for these phases.
They appear to be too low to self-stabilize the clumps against their high 
internal turbulence. Since these values clearly fall below the Roche density 
of $>$10$^7$ cm$^{-3}$ at the position of the CND, the authors conclude 
that the bulk of the material in the CND does not consist of stable structure 
and must be rather transient. 
Densities of $\sim$10$^7$ cm$^{-3}$ are indicated for CND molecular 
clumps through interferometric HCN and HCO$^+$ measurements by Christopher et al. (2005).

\subsection{Variability}
\label{variability}
In a comprehensive statistical approach  and using the complete 
VLT NACO Ks-band data between 2002 and 2010, Witzel et al. (2012)
could give a self-consistent statistical description of the NIR variability
of SgrA*.
Witzel et al. (2012) could not confirm the earlier claimed two 
states of variability (Dodds-Eden et al.  2011).
We find the NIR variability to be a single-state process
forming a power-law distribution of the flux density.
This distribution even offers an 
explanation for the claimed X-ray flare that probably occurred 400 years ago 
(Revnivtsev et al. 2004, Sunyaev \& Churazov 1998, Terrier et al. 2010)
as an extreme 
value of our flare statistics without the need for an extraordinary event.
However, it cannot be fully excluded that dusty objects, 
like the DSO/G2, the X3 and X7 sources, or smaller cloudlets 
(Gillessen 2012, Eckart et al. 2013, Muzic et al. 2010) may be responsible for
rare and sporadic off state accretion events.
In bright flares - as they are expected if the DSO/G2 leads to an enhanced 
accretion during or past its SgrA* flyby - QPOs (quasi periodic oscillations) 
may occur that probe the last stable orbit
region of SgrA* and allow to constrain the BH spin 
(Zamaninasab et al. 2010; Eckart et al.  2006).

These phenomena are observed in nearby galactic nuclei:
Temporary QPO features have also been detected in 
Seyfert~1 nucleus of RE J1034+396 (Middleton, Uttley, \& Done 2011).
Cenko et al. (2012) report on a short-lived, luminous flare from the nuclear region of a star-forming galaxy
that may be the result of an interaction between a stellar object and a SMBH.
The conditions under which these phenomena occur can probably be studied best 
at the center of the Milky Way.

There are various approaches to physically model
and understand the steady state variability of SgrA*. 
Eckart et al. (2012) use the total NIR/X-ray flux densities and 
a Synchrotron/Self-Compton model with 
adiabatic expansion to match the radio.
The adiabatic time-lag is measured best with the large frequency 
separation offered by LABOCA/SABOCA (345 \& 850 GHz).
The typical adiabatic expansion speeds of $\le0.1c$ are too slow for material 
to leave the immediate vicinity of SgrA*. 
The expansion must either take place in an orbiting disk or in a jet 
component with a higher bulk velocity.
Larger scales, i.e. separations of $\le$0.5'' i.e. within the Bondi sphere -
are usually described by magneto-hydrodynamical (MHD) models 
(Moscibrodzka et al. 2009, Dexter et al. 2010, Shcherbakov \& Baganoff 2010).
The mid-plane of these models can be described using a 3D relativistic disk code
(e.g. Valencia-S., Bursa et al.  2012; capable of investigating the effect of radiative transport and polarization). 
Indications for a 3~pc scale jet recently outlined by Yusef-Zadeh et al. (2012) are in 
agreement with the overall orientation of the SMBH and the corresponding spin-axis
as derived by Zamaninasab et al. (2011).

\section{Conclusion}
\label{conclusion}
Based on the work of Contini (2011) and Ho (2008), the Galactic 
Center can be regarded as a LLAGN in a faint or even off-state.
Based on recent work in the radio, infrared and X-ray regime,
the center of the Milky Way has all the necessary building blocks
that may be required for a bona fide LLAGN.
The analysis of the central stellar cluster indicates the
presence of young stars (e.g. Buchholz et al. 2009, Sch\"odel et al. 2009, 2010;
see also Arches and Quintuplet clusters). 
Both the NACO VLT and NIRC2 Keck data support the 
identification of the DSO/G2 with a 2$\mu$m continuum source.
This suggests that we see photospheric emission from dusty star (details in Eckart et al. 2013).
The uncertainties between the K- and L'-band measurements obtained by the
VLT and Keck suggest that the orbit of the DSO is currently not well defined.
Dusty sources that may be linked to young and recently formed stars 
are also present in the central stellar cluster 
(Eckart et al. 2013 and references there in).
Combined with the sub-mm observations that reveal dusty molecular cloud 
complexes, this shows that there is the potential for ongoing star 
formation in the LLAGN at the center of the Milky Way.
There are three main points that can be concluded:
i) There is an indication that the overall efficiency for accretion 
events in LLAGN is low (Ho 2008) which is consistent with  
the SgrA* NIR variability statistics that can explain the 
X-ray flare about 400 years ago (Witzel et al. 2012); 
ii) There is an observational indication of an accretion wind 
associated with the SMBH SgrA* (e.g. Muzic et al. 2007, 2010, Zamaninasab 2011); 
iii) Sporadic accretion events can be expected in the near future due
to the DSO flyby (Gillessen et al. 2012, Eckart et al. 2013).
This scenario supports the assumption that the Milky Way may 
in fact harbor a LLAGN at its center.    
\\
\\
{\bf Acknowledgements:}
{\small
This work was supported in part by the Deutsche Forschungsgemeinschaft
(DFG) via the Cologne Bonn Graduate School (BCGS), 
and via grant SFB 956, as well as by 
the Max Planck Society and the University of Cologne through 
the International Max Planck Research School (IMPRS) for Astronomy and 
Astrophysics.
This research has made use of the Keck Observatory Archive (KOA), which is 
operated by the W. M. Keck Observatory and the NASA Exoplanet Science Institute (NExScI), 
under contract with the National Aeronautics and Space Administration. 
The datasets used have the following PIs: J.Lu, M.R.Morris \& T.Soifer.
We had fruitful discussions
with members of the European Union funded COST Action MP0905: Black
Holes in a violent Universe and the
COST Action MP1104:
Polarization as a tool to study the Solar System and beyond.
We received funding from the 
European Union Seventh Framework Programme (FP7/2007-2013) 
under grant agreement No.312789.
}
\\
\\

\end{document}